\documentclass[10pt]{iopart}
\begin{document}
\title[Hamiltonians for curves]{Hamiltonians for curves}
\author{R Capovilla\dag, C Chryssomalakos\ddag and
J Guven\ddag}
\address{\dag\
Departamento de F\'{\i}sica,
Centro de Investigaci\'on y de Estudios Avanzados del IPN,
 Apdo. Postal 14-740, 07000 M\'exico DF, MEXICO}
\address{ \ddag\ Instituto de Ciencias Nucleares,
Universidad Nacional Aut\'onoma de M\'exico,
 Apdo. Postal 70-543, 04510 M\'exico DF, MEXICO}

\begin{abstract}
We examine the equilibrium conditions of a curve in space when 
a local energy penalty is 
associated with its extrinsic geometrical state characterized by its  
curvature and torsion.
To do this we tailor the theory of deformations to the Frenet-Serret 
frame of the curve. The Euler-Lagrange equations describing 
equilibrium are obtained; 
Noether's theorem is exploited to identify the constants 
of integration of these equations as the Casimirs of the euclidean group
in
three dimensions. 
While this system appears not to be integrable in 
general, it {\it is} in various limits of interest. 
Let the energy density be given as some function of the  curvature and
torsion, 
$f(\kappa,\tau)$.
If $f$ is a linear function of either of its arguments 
but otherwise 
arbitrary, we claim that
the first integral associated with rotational invariance permits the 
torsion $\tau$ to be expressed as the 
solution of an algebraic equation in terms of the bending curvature,
$\kappa$. The first integral associated 
with translational invariance can then be cast as a 
quadrature for $\kappa$ or for $\tau$.
\end{abstract}


\section{Introduction}

Consider a curve in space. Suppose that the
curve is sufficiently smooth so that the Frenet-Serret
frame adapted to it is defined. The curvature 
$\kappa$ and the torsion $\tau$ then provide a complete
characterization of the curve; once they are known,
it can be reconstructed up to euclidean motions.
In this paper we examine local 
reparametrization invariant hamiltonians for 
curves of the form 
\begin{equation}
H = \int ds \; f ( \kappa , \tau )\,,
\label{eq:kat}
\end{equation}
where $s$ denotes arclength, and  $f$ is any scalar under
reparametrizations. 

Such hamiltonians play a role both in the static and in the kinematic
description of curves.
In the former, we interpret the hamiltonian as the energy of the 
physical system; 
in particular, an energy of the form $f = \kappa^2 $
penalizing bending models the stiffness of a polymer \cite{Doi.Edw:88,Gol.Lan:95},  
and it has been used to model the elastic properties of DNA 
(see {\it e.g.} \cite{Bou.Mez:98,Mar.Sig:95}). 
When the energy depends only on the
curvature, as
Bernoulli and Euler both knew, the equilibrium conditions are integrable
\cite{Gia.Hil:96}: the torsion is always a function of
$\kappa$, and $\kappa$ satisfies a quadrature.
Our focus will be on the general case (\ref{eq:kat}). 
Whereas a linear dependence on
$\tau$, associated with a constraint on the total torsion, has been 
considered \cite{Ive.Sin:99}, very little appears 
to be known beyond that \cite{Wil:00}. 
At the very least, one is
interested in a quadratic dependence on both
$\kappa$ and $\tau$, the second order terms in a Taylor expansion  of $f$.
Such terms  appear in the hamiltonian describing chiral polymers 
\cite{Kam:02}.  

An  additional motivation for studying these hamiltonians is the role they
play in the connection between the motion of curves and integrable systems:
the equation describing the evolution of some function of the 
curvature and the torsion with respect to certain length preserving 
vector fields 
coincides with the nonlinear Schr\"odinger equation; 
other functions give other known integrable equations.
These curve motions appear in a number of contexts:
vortex filaments and patches in fluids \cite{Has:72,Lam:76,Gol.Pet:92},  
classical magnetic spin chains \cite{Lak:77,Bal:82}, interface
dynamics \cite{Gol.Pet:91}, etc. 
Specific hamiltonians of the form
(\ref{eq:kat}) emerge  as conserved quantities under these
motions  \cite{Lan.Per:94,Lan:99}. 

Our strategy in the study of the hamiltonians (\ref{eq:kat}) 
will be to exploit Noether's theorem 
to identify the equilibrium conditions as conservation
laws associated with the euclidean invariance of the
energy. These conservation laws in turn
permit us to identify first integrals of the
equilibrium conditions. Remarkably, in certain 
cases, these integrals can be combined to provide
a quadrature for either $\kappa$ or $\tau$. The
constants of integration are the Casimir invariants
of the euclidean group. We show that, in addition to
the well known case of a pure bending hamiltonian,
a pure torsion hamiltonian also leads to integrable equilibrium conditions.
This is surprising because, in contrast to the curvature which depends 
on two derivatives of the embedding function for the curve,  
the torsion depends on three.
The Euler-Lagrange equations which result  
involve six derivatives; as such one would not expect them to be 
tractable. The torsion is determined by a quadrature. For a 
polynomial $f$, the potential appearing in this quadrature is a rational function.
We identify other hamiltonians
with a joint dependence on $\kappa$ and $\tau$ which are 
reducible to a quadrature. In general, unfortunately,
it does not appear to be possible to identify a quadrature. 
The integrals of the motion
can, however, be used to reduce the equilibrium conditions to the motion
of a fictitious particle in two dimensions. In any case,
this reduction should be helpful for studying these systems.

For most physically realistic materials the local arclength will be constant.
This is because there will be a large energy penalty associated with 
stretching the curve. Suppose that an arbitrary deformation is decomposed into
tangential and normal parts. The constraint on the arclength  
can then be phrased in terms of the corresponding response of the
tangential deformation to its normal counterpart.
However, as we will discuss below, 
a tangential deformation is a reparametrization so that the corresponding
change in the hamiltonian can always absorbed in a divergence; as such, it
cannot affect the Euler-Lagrange equations. 
Thus, whether or not we decide to implement a constraint on arclength, 
the equations themselves describing equilibrium are unchanged.

The paper is organized as follows.
In Sect. 2, we begin by giving a self-contained account of the theory
of deformations of a curve tailored to the
Frenet-Serret frame. 
In distinction to earlier work, it is not necessary to 
implement the constraint associated with locally arclength preserving
deformations. We obtain directly simple expressions
for the variation of the curvature and torsion.
In Sect. 3, we analyze the consequences of the invariance
of the hamiltonian under reparametrizations as well as under euclidean
motions. We show how to obtain expressions for the internal forces
and torques on any segment of a curve and their relationship with
the equilibrium conditions.
Systems which depend at most on the curvature $\kappa$ is
the subject of Sect. 4. This is extended to systems that depend on the
torsion $\tau$ in Sect. 5. In Sect. 6, systems that depend both
on curvature and torsion are considered. In Sect. 7 we briefly consider
perturbations of the equilibrium conditions.
Sect. 8  relates some of the results of this paper to
recursion schemes that appear in the kinematics of curves.

\section{Curve deformations}

In this section we describe the geometry of embedded 
curves in three dimensional space in terms of the 
Frenet-Serret basis for the curve, and 
the effect of a small deformation of the curve on its geometry.

Consider a curve in space described by the embedding 
$ {\bf x} ={\bf X}(s )$, where ${\bf X} = (X^1, \, X^2, \, X^3)$.
The unit tangent to the curve is given by ${\bf  t}= {\bf X}{}'$, 
where prime denotes a derivative with respect to arclength $s$.  
Clearly, the `acceleration' ${\bf  t}'$ is orthogonal to ${\bf  t}$.
However, ${\bf  t}''$ is not.  
The classical Frenet-Serret equations 
\begin{eqnarray}
{\bf t}' &=&  \kappa {\bf  n}_1 \,, 
\nonumber
\\
{\bf  n}_1{}' &=& - \kappa {\bf  t} + \tau  {\bf  n}_2 \,, 
\label{eq:frenet}
 \\
{\bf  n}_2{}' &=& - \tau {\bf  n}_1 \,, \nonumber
\end{eqnarray}
describe the construction of an  orthonormal basis 
$\{{\bf  t} , {\bf  n}_1,{\bf  n}_2 \} $ along the curve. We choose an
orientation with ${\bf  n}_2 =  {\bf  t} \times {\bf  n}_1 $.
$\kappa$ and  $\tau$ are respectively  the geodesic curvature and torsion.
The fundamental theorem for curves tells us that the  
Frenet-Serret curvatures $\kappa$ and $\tau$ determine the curve up to
rigid motions \cite{Spi:79}. The actual curve can always
be reconstructed from its curvatures. 
Thus, they provide a natural set of auxiliary variables. 
Any local geometrical scalar 
defined along the curve can in principle 
always  be expressed as a function of the curvatures and their derivatives.

We now analyze the change in the geometry of
the curve due to an infinitesimal deformation of its embedding in space,
${\bf  X} (s ) \to {\bf  X} (s ) + \delta {\bf  X} (s ) $.
Let us first decompose the deformation into its tangential and normal parts
with respect to the basis $\{{\bf  t} , {\bf  n}_1 , {\bf n}_2 \}$,
\begin{equation}
\delta {\bf  X}  = \Psi_\parallel \; {\bf  t}  + \Psi_1 \; {\bf  n}_1 +
\Psi_2 {\bf n}_2\,.
\end{equation}
This is a convenient strategy when one is interested in the
variation of reparametrization invariant geometrical quantities.
Tangential deformations are reparametrizations
of the curve. We will use the following two facts:
the tangential deformations of any scalar $f$ and of the infinitesimal
arclength are given by
\begin{equation}
\delta_\parallel f =   \Psi_\parallel \; f' 
\,, 
\qquad 
\qquad
\qquad
\delta_\parallel ds  = ds \; \Psi_\parallel{}'
\,.
\label{eq:st}
\end{equation}
Now consider the normal part of the deformation.
A normal deformation of $ds$ is 
\begin{equation}
\delta_\perp ds  = - ds  \; \kappa \; \Psi_1 
\,.
\label{eq:sn}
\end{equation}
This result implies that, for any scalar $f$,
\begin{equation}
\delta_\perp ( f' ) =   \kappa \; f' \, \Psi_1  + 
( \delta_\perp  f )'
 \,.
\label{ds}
\end{equation}
In particular, for the three scalar functions ${\bf X}$, this implies
\begin{equation}
\delta_\perp{\bf  t} =
\kappa \Psi_1 {\bf  t} 
+ \left(\Psi_1 {\bf  n}_1 + \Psi_2 {\bf  n}_2 \right)' 
\, .
\label{dperpt}
\end{equation}
We now use the Frenet-Serret equations to cast $\delta_\perp{\bf  t}$ 
as a normal vector (a unit vector and its variation are orthogonal):
\begin{equation}
\delta_\perp{\bf  t} = 
(\Psi_1{}' - \tau \Psi_2){\bf  n}_1 
+ (\Psi_2{}' + \tau \Psi_1) {\bf  n}_2
\, .
\label{eq:xpp}
\end{equation}
Similarly we have 
\begin{equation}
\delta_\perp{\bf  t}' =    
\kappa \Psi_1 {\bf  t}' + (\delta_\perp{\bf  t} )'
\, ;
\label{dperptp}
\end{equation}
using the Frenet-Serret equations and~(\ref{eq:xpp}) we obtain 
\begin{eqnarray}
\delta_\perp{\bf  t}' 
&=&
- \kappa \left( \Psi_1{}' 
- \tau \Psi_2 \right)  {\bf  t} 
\nonumber \\
& &
{}+ \left[ \Psi_1{}'' +(\kappa^2 -\tau^2) \Psi_1 
- 2 \tau \Psi_2{}' - \tau{}' \Psi_2 \right] {\bf n}_1
\nonumber \\
& & 
{}+\left( 2 \tau \Psi_1{}' 
+ \tau' \Psi_1  +\Psi_2{}'' 
-\tau^2 \, \Psi_2 \right) {\bf n}_2 
\, .
\label{eq:xxpp}
\end{eqnarray}

We are now in a position to evaluate the normal variations of the
two Frenet-Serret curvatures. To evaluate $\delta_\perp \kappa$,
we take a variation of the first of~(\ref{eq:frenet}): 
$ \delta_\perp{\bf  t}' =  (\delta_\perp \kappa ) {\bf  n}_1 
+ \kappa \delta_\perp {\bf  n}_1$.
Dotting with ${\bf n}_1$ we obtain
$\delta_\perp \kappa = {\bf  n}_1 \cdot \delta_\perp{\bf  t}' \,.
$
From (\ref{eq:xxpp}) we read off that
\begin{equation}
\delta_\perp \kappa =  
\Psi_1{}'' 
+ \left( \kappa^2 - \tau^2 \right) \Psi_1
- 2 \tau \Psi_2{}' - \tau' \Psi_2   
\, .
\label{eq:perpkappa}
\end{equation}
For a planar curve, $\tau=0$ and
$\delta_\perp \kappa = 
 \Psi_1{}'' + \kappa^2 \Psi_1$. 
Deformations lifting the curve off the plane do not affect
the value of $\kappa$ to first order in the deformation. 

To evaluate $\delta_\perp \tau$, we take a variation of the
FS equation for ${\bf n}_1'$  
and dot with ${\bf  n}_{
2}$. We have 
\begin{equation}
\delta_\perp \tau = 
\kappa {\bf  n}_2 \cdot \delta_\perp {\bf t} + {\bf  n}_2 \cdot
\delta_\perp {\bf  n}_1{}'\,.
\end{equation}
We now rewrite the second term on the right hand side as
\begin{eqnarray}
{\bf  n}_2 \cdot
\delta_\perp {\bf  n}_1{}'  
&=&
\kappa \Psi_1 ( {\bf  n}_2 \cdot {\bf  n}_1{}' ) 
+ {\bf  n}_2 \cdot ( \delta_\perp {\bf  n}_1 )'  
\nonumber \\
&=&
\kappa \tau \Psi_1  + 
({\bf  n}_2 \cdot \delta_\perp {\bf  n}_1 )'
\nonumber \\
&=&
\kappa \tau \Psi_1 
+ \left[ {1 \over \kappa}  {\bf  n}_2 \cdot \delta_\perp{\bf  t}'
\right]'
\, ,
\end{eqnarray}
where we have applied~(\ref{ds}) to ${\bf n}_1'$, 
and we have used the FS equations for both ${\bf n}_1'$ 
and ${\bf n}_2'$.
Substituting for ${\bf  n}_2 \cdot 
\delta_\perp{\bf  t}$ and ${\bf  n}_2 \cdot \delta_\perp{\bf  t}'$, 
we obtain
\begin{equation}
\delta_\perp \tau = 
\kappa (\Psi_2{}'  + 2 \tau \Psi_1) +  
\left\{ {1\over \kappa} \left[ 
2 \tau \Psi_1{}' + \tau' \Psi_1
+ \Psi_2{}''   
- \tau^2 \Psi_2 \right]\right\} '
\, .
\label{eq:varkappa2b}
\end{equation}
For an initially planar curve,
$\delta_\perp  \tau = 
\kappa \Psi_2{}'  +
(\Psi_2{}''/\kappa)'$. Only the deformation along the direction normal 
to the 
plane contributes.
Suppose that $\delta_\perp \tau=0$, then the equation 
$\kappa\Psi_2{}'  +(\Psi_2{}''/\kappa)' =0$
should not admit any solutions other than those which correspond to an 
euclidean motion of the 
planar curve. To show this we note that, for a planar curve, $\kappa
= \Theta'$, where
$\Theta$ is the angle which the tangent makes with, say, the $x$-axis.
Then this equation
can be recast as $\partial_{\Theta}^2 \Psi'_{2} 
+ \Psi_2' = 0$,
with independent solutions, 
$\Psi_2= \sin\Theta,\cos\Theta$ and $\Psi_2$ a constant ---
a rotation about $x$, $y$ and a translation. If $\delta_\perp\tau$ 
is constant, the 
only solution is the helix $\Psi_2 \propto 
\Theta$ generated by the planar curve.

For completeness, let us note that the normals vary according to
\begin{eqnarray}
\delta_\perp {\bf n}_1 &=& 
- (\Psi_1{}' - \tau \Psi_2){\bf  t}
+  {1\over \kappa} \left( 
 \Psi_2{}'' - \tau^2 \Psi_2 
 + 2 \tau \Psi_1{}' + \tau' \Psi_1
\right) {\bf n}_2\,,
\label{n1} \\
\delta_\perp {\bf n}_2 &=& 
- (\Psi_2{}' + \tau
\Psi_1){\bf  t} -
  {1\over \kappa} \left( 
 \Psi_2{}'' - \tau^2 \Psi_2 
 + 2 \tau \Psi_1{}' + \tau' \Psi_1
\right) {\bf n}_1\,.\label{n2}
\end{eqnarray}

So far we have considered arbitrary deformations of the curve. There
are special deformations that will be of interest in the following.
In particular a deformation ${\bf Y} = \delta {\bf X} $ preserves locally
the arclength if
$\delta_{{\bf Y}} ds = 0$. In terms of the components 
this
translates to
\begin{equation}
{\bf t} \cdot {\bf Y}' = Y_\parallel{}' - \kappa Y_1 = 0
\, ,
\label{eq:lap}
\end{equation}
which implies that there exists a (non-unique) vector ${\bf Z}$ such
that
\begin{equation}
{\bf t}  \times 
{\bf Z} =  {\bf Y}' 
\, .
\label{eq:lap1}
\end{equation}
This is the starting point of the filament model recursion 
scheme \cite{Lan:99}, where a family of locally arclength preserving
vector fields 
$\{  {\bf Y}_{(n)} \}$ is defined by
$
{\bf t}  \times 
{\bf Y}_{(n)} =  {\bf Y}'_{(n-1)} \,,
$
with ${\bf Y}_{(0)} = - {\bf t}$.
We will have something to say about this  below in Sect.~\ref{FM}.

Let us also note that if $ (\delta_\perp {\bf X} )' \cdot {\bf n}_i = 0$,
or in components, $ \Psi_1{}' - \tau \Psi_2 = 0$ and $ 
\Psi_2{}' + \tau \Psi_1 = 0 $, then 
$ \delta_\perp {\bf n}_i = 0$ and $\delta_\perp {\bf t} = 0$; the
Frenet-Serret basis is left unchanged by this type of deformation.  
\section{Invariance and symmetry}
The hamiltonian $H$ for the curve depends locally on the geometry, and it
possesses various symmetries, both local and global. The local symmetry 
is reparametrization invariance, and it restricts severely the
form of $H$.
The global symmetries are euclidean motions: translations
and rotations. They give rise to
conservation laws.
\subsection{Reparametrization invariance}
The hamiltonian $H$ is, in general,  a sum of terms, 
$H= H_1 + H_2 + \cdots$, each 
of which is invariant
under reparametrizations of the curve. This results in the form  
\begin{equation}
H = \int ds \; f ( \kappa , \tau, \kappa' , \tau' , \cdots ) 
\,,
\end{equation}
where $f$ is a scalar under reparametrizations, constructed out
of the geometrical quantities that characterize the
curve:  the two curvatures, and their derivatives.
The lowest order in derivatives of the embedding functions non-trivial
geometrical models depend 
only on the scalar $\kappa$. A simple model that 
penalizes bending of the curve is
$f_1 = {\bf t}' \cdot {\bf t}' =
\kappa^2$. 
At the next order, a dependence on $\tau$ as well as 
$\kappa'$ is admitted. 
A term in  $f$ of the form $\tau^2 $ penalizes the torsion of the
curve; 
one of the form $(\kappa')^2$ is a higher order 
differential bending energy and will not be 
considered. We note that the natural hamiltonians 
$f_2 \equiv {\bf n}_1'\cdot {\bf n}_1'$ and 
$f_3 \equiv {\bf n}_2'\cdot {\bf n}_2'$ are given by
$f_2=\kappa^2 + \tau^2$ and $\tau^2$. Whereas $f_1$ has been 
considered in considerable
detail \cite{Lan.Sin:84,Bry.Gri:86}, neither $f_2$ nor $f_3$ 
appear to
have been 
considered.

The assumption that $f$ is a scalar under reparametrizations
implies that
the first variation of 
the energy can always be written as
\begin{equation}
\delta H  = \int ds \; {\cal E}_i \; \Psi_i +
 \int ds \; {\cal Q}' \,,
\label{eq:var}
\end{equation}
where ${\cal E}_i$ denotes the normal projection of the 
Euler-Lagrange derivative of $f$, and  ${\cal Q}$ is the Noether
charge ($i,j, \cdot = 1,2$).
The specific form of the first term follows from the fact that
the tangential variation contributes only in a divergence. 
Indeed, using~(\ref{eq:st}),
the tangential part of the variation of the energy is always
a total derivative
\begin{equation}
\delta_\parallel H
= \int ds  \;  (  f \; \Psi_\parallel )' \,.
\label{eq:varpa}
\end{equation}
This implies that the Noether charge ${\cal Q}$, 
as a linear
differential
operator which operates on the deformation 
$\delta {\bf X}$, is of the form
\begin{equation}
{\cal Q} = f \; \Psi_\parallel + 
{\cal Q}_{(0)}{}^i \; \Psi_i 
+ {\cal Q}_{(1)}{}^i \; \Psi_i{}' 
+  \cdots\,,
\label{eq:charge}
\end{equation}
where, to construct the ${\cal Q}_{(n)}{}^i$, we use integration
by parts to collect in a total derivative the normal deformations
$\Psi_i$ and their derivatives.
\subsection{Translational invariance}
The hamiltonian $H$ is also 
invariant under the rigid euclidean motions: translations and rotations.
Noether's theorem can then be exploited to 
determine the conditions of static equilibrium. 
Under an infinitesimal constant translation, $\delta {\bf  X} = {\bf  e}$, 
the energy associated with $H$ stored on the segment 
$AB$ (labeled by its endpoints) changes by an amount 
\begin{equation}
\delta H_{AB} = {\bf   e}\cdot 
\int_{A}^B ds \;  \left[ {\cal E}^i \; {\bf  n}_i - {\bf  F}'\right]
\,.
\end{equation}
Here we introduce the spatial vector ${\bf F}$ with
\begin{equation}
{\cal Q}= - {\bf e} \cdot {\bf F}\,,
\end{equation}
and it is constructed by specializing Eq.~(\ref{eq:charge}) to the case
of a constant deformation.

With no external forces acting so that  ${\cal E}_i=0$, 
we may identify ${\bf F}$ as the internal force pulling or pushing 
at a point on the  curve segment.
(With our conventions, ${\bf F}$ is the force from the part with lower
to the one with higher value of $s$.)
In general, this force will not be tangential.
Translational invariance implies  $\delta H_{AB}=0$.
Because the endpoints $A$ and $B$ are arbitrary we deduce  the
local balance of forces
\begin{equation}
{\cal E}^i \; {\bf  n}_i - {\bf  F}' = 0
\, 
\label{eq:Eq}
\end{equation} 
In equilibrium, 
this implies the 
conservation law, ${\bf  F}'=0$, {\it i.e.}  ${\bf  F}$ 
is a constant vector along the curve. We thus associate  a spatial vector 
with each curve, be it open or closed. The squared magnitude of this 
vector, $F^2$,  is the first Casimir of the euclidean group.
The direction along which ${\bf F}$ points is often indicated 
by the symmetry of the configuration.

We have come up with three 
equations of static equilibrium, whereas we 
only possess two independent Euler-Lagrange equations. 
One of the former must therefore be a kinematical statement,
or Bianchi identity, associated with the 
reparametrization invariance of $H$. 
Let us examine the three independent projections 
of the equilibrium conditions, Eq.~(\ref{eq:Eq}):
we decompose ${\bf  F}$ into parts tangential and 
normal to the curve,
\begin{equation}
{\bf  F} = F_{||} \; {\bf  t} + F_1 \; {\bf  n}_1 + F_2 \; {\bf n}_2 \,. 
\end{equation}
Then Eq.~(\ref{eq:Eq}) is equivalent to the three equations
\begin{eqnarray}
F_{||}' &-& \kappa \; F_1  
= 
0\,, 
\nonumber \\
F_1{}' &+& \kappa \; F_{||} - \tau \; F_2 
=  
{\cal E}_1\,, 
\\
F_2{}' &+& \tau \; F_1 
=  
{\cal E}_2
\,. 
\nonumber
\label{eq:3fs}
\end{eqnarray}
Comparison with~(\ref{eq:lap}) shows that  ${\bf F}$,
seen as a deformation of the curve, 
preserves locally arclength, as expected.
The first condition is independent of the 
Euler-Lagrange equations. It is the promised Bianchi identity
associated
with 
the reparametrization invariance of $H$. 
Note that these equations could also 
have been derived directly by considering the balance of forces,
as in~\cite{Arr.Cap.Chr.Guv:02}.

There is a non-trivial integrability condition on closed curves
associated with the conservation law (\ref{eq:Eq}). Taking its projection
onto ${\bf X}$ in equilibrium, 
we can immediately deduce 
that 
\begin{equation}
\oint ds\; F_{||} = 0
\, .
\label{eq:int}
\end{equation} 
on any closed loop. We will comment below in Sect. 3.4. 
on its geometrical origin.
\subsection{Rotational invariance}
Under an infinitesimal rotation $\delta {\bf  X} = 
{\bf  \Omega}\times {\bf  X}$,  we have that the energy $H$
on a segment $AB$ of the curve changes by 
\begin{equation}
\delta H_{AB} = {\bf  \Omega}
\cdot 
\int_A^B ds \; \left[
{\cal E}^i \; {\bf
n}_i \times {\bf  X}  - {\bf M}{}'\right]\,,
\end{equation}
where the spatial vector ${\bf M}$ is defined by
\begin{equation}
{\cal Q}= - {\bf \Omega} \cdot {\bf M}\,, 
\end{equation}
and it is obtained from Eq.~(\ref{eq:charge}) by 
specializing it to the case of a constant rotation.
We identify ${\bf M}$ as the torque with respect to the origin  acting at
a 
point on the curve segment. Rotational invariance of $H_{AB}$ implies
\begin{equation}
{\bf M}' = {\cal E}^i \; {\bf
n}_i \times {\bf  X} \,.
\end{equation}
We decompose ${\bf M}$ into the sum of the couple of ${\bf F}$ about 
the origin, plus an intrinsic, translationally invariant part,
\begin{equation}
{\bf M} = {\bf X} \times {\bf  F}  + {\bf T}\,.
\end{equation}
Then the differential torque ${\bf T}$  satisfies 
\begin{equation}
{\bf T}{}' =  {\bf  F} \times {\bf t}\,. 
\label{eq:TFt}
\end{equation} 
We emphasize that this equation does not depend on whether
the curve is in equilibrium or not.
Whereas ${\bf M}$ is conserved in equilibrium, 
neither ${\bf X}\times {\bf F}$
nor ${\bf T}$ is. It is clear, however, that the projection of 
${\bf T}$ onto ${\bf F}$, 
the second Casimir of the  euclidean group, is 
conserved
\begin{equation}
J = {\bf T}\cdot {\bf \hat{F}}\,
\label{eq:tau}
\end{equation}
where ${\bf \hat{F}} = {\bf F} / F $, and for future convenience, we
choose 
to fold $|{\bf F}|$ into the definition.

As we did earlier for ${\bf F}$,
we can also decompose ${\bf T}$ into parts tangential and normal to the
curve,
\begin{equation}
{\bf T} =  T_\parallel \; {\bf  t} + T_1 \; {\bf  n}_1 + T_2 \; {\bf  n}_2 
\,.
\end{equation}
Then Eq.~(\ref{eq:TFt}) is equivalent to 
\begin{eqnarray}
T_\parallel'  &-& \kappa  T_1  
= 
0\,, 
\nonumber\\
T_1{}' &-& \tau  T_2 + \kappa T_\parallel 
=  
F_2\,, 
\\
T_2 {}' &+& \tau  T_1 
= 
- F_1 
\, ,
\nonumber
\end{eqnarray}
where we are using  the convention ${\bf t} \cdot {\bf n}_1 \times {\bf
n}_2
= 1 $. The first equation, as was the case for ${\bf F}$ in Eq. 
(\ref{eq:3fs}), is also
valid off-shell, and it says that ${\bf T}$, seen as a deformation 
of the curve preserves locally arclength. 

If a curve is deformed along ${\bf T}$, {\it i.e.} 
$\delta {\bf X} = {\bf T}$, we find that the corresponding
variation of the
curvature and torsion satisfy
\begin{equation}
\delta_{\bf T} \kappa = {\cal E}_2 
\,, 
\qquad 
\qquad 
\qquad 
\delta_{\bf T} \tau = - \left( { {\cal E}_1 \over \kappa} \right)'  
\,.
\end{equation}
Therefore, in equilibrium, not only deformations along  ${\bf F}$, 
${\bf M}$ corrrespond to rigid motions which leave the geometry
unchanged, as expected, but also deformations along ${\bf T}$. 
\subsection{Adapted cylindrical coordinates}
The two 
conserved vectors 
${\bf F}$ and ${\bf M}$ together single out a  
cylindrical polar coordinate system $\{\rho, \theta, z \}$
\cite{Lan.Lif:70,Lan.Sin:84,Lan.Sin:84a}. Suppose ${\bf F}\ne 0$.
We align our coordinate system such that 
$\hat {\bf F}$ points along the positive $z$ direction.
We next perform a translation orthogonal to ${\bf F}$ 
so that $\hat{\bf M}$ is rotated into $\hat {\bf F}$.
Then we have that
\begin{equation}
{\bf T} = J \,   {\bf \hat{F}} + {\bf F} \times {\bf  X}  
\, .
\label{eq:coord}
\end{equation}

Alternatively, we can arrive at this expression by integrating Eq.
(\ref{eq:TFt}), 
\begin{equation}
\int ds \; {\bf T}' = \int ds \; ( {\bf F} \times {\bf X} )' - 
\int ds \; {\bf F}' \times {\bf t}\,,  
\end{equation}
and noting that at equilibrium the second term vanishes, so that
${\bf T}$ and ${\bf F} \times {\bf X}$ differ by a constant vector.
Then contraction with ${\bf \hat{F}}$  reproduces Eq.
(\ref{eq:coord}),
up to a constant translation.

The modulus 
\begin{equation}
T^2 =  J^2  + \rho^2 F^2
\label{t1}
\end{equation}
determines the cylindrical radius in the adapted system
in terms of $T^2$ and the two
Casimirs. Typically, $T^2$ will be some function of
$\kappa$ and $\tau$ and their derivatives. 
To complete the construction, we describe the tangent vector in these 
coordinates,
${\bf t} = (\rho',\theta',z')\,.$
Thus, the projection ${\bf t}\cdot \hat {\bf F}$ 
determines $z$ as a quadrature 
\begin{equation}
z'  = { F_\parallel \over F}\,.
\label{t2}
\end{equation}
This provides the promised geometrical interpretation of the 
global conservation law  (\ref{eq:int}).
Similarly, the projection ${\bf t}\cdot \hat {\bf T}$ 
determines $\theta$ as a quadrature,
\begin{equation}
F \rho^2 \theta' =  J z'  - T_\parallel \,.
\label{t3}
\end{equation}
The expressions (\ref{t1}), (\ref{t2}) and (\ref{t3}) 
are 
related by the normalization condition 
$\rho'{}^2 + \rho^2 \theta'{}^2 + z'{}^2=1$. 
This is not immediately apparent. It can be shown by
taking a derivative
of Eq. (\ref{t1}) and squaring, 
together with the squares of Eqs. (\ref{t2}), (\ref{t3}),
as long as $\rho$ does not vanish.
\section{Bending}
Let us consider an hamiltonian that
depends at most on the  curvature,
\[
H = \int ds \;  f( \kappa )\,,
\]
where $f$ is any local function of its argument.
We find that under an arbitrary deformation of the curve 
we have, with $f_\kappa = \partial f / \partial \kappa $,
\begin{eqnarray}
\delta H 
&=& \int ds \, \left\{ f_\kappa \delta_\perp \kappa - f \kappa \Psi_1
\right\}
+ \int ds \,  (f \Psi_\parallel)' \nonumber\\
&=& \int ds  \, \Big[  f_\kappa{}'' + (\kappa^2 - \tau^2) f_\kappa 
- \kappa f\Big]\Psi_1 
 + \int ds \, \Big[ 
 (2 \tau f_\kappa)' - f_\kappa \tau' \Big] \Psi_2
\nonumber\\
&&+ \int ds \; \left[ f_\kappa\Psi_1{}' - f_\kappa{}' \Psi_1
+ f \Psi_\parallel - 2 \tau f_\kappa \Psi_2\right]'\,,
\label{eq:delFS}
\end{eqnarray}
where we have used the expression (\ref{eq:perpkappa}) for 
$\delta_\perp \kappa $. By comparison with Eq. (\ref{eq:var}),
we immediately read off the Euler-Lagrange 
equations ${\cal E}_i=0$, where
\begin{eqnarray}
{\cal E}_1  
&=&
f_\kappa{}'' + (\kappa^2 - \tau^2) f_\kappa - \kappa f 
\label{eq:E1}
\\
{\cal E}_2 
&=&  
(2 \tau f_\kappa )' - \tau' f_\kappa 
\, .
\label{eq:E2}
\end{eqnarray}
We see that $\tau$ contributes
to the ``driving force"  for $\kappa$ in the first
equation.
Integrating the second gives
\begin{equation}
f_\kappa{}^2  \tau = {\rm constant}\,,
\label{f12}
\end{equation}
which determines $\tau$ as a function of $\kappa$. 
We show below that the constant appearing here is $J$ as defined by 
Eq.(\ref{eq:tau}). 
We can substitute into Eq.(\ref{eq:E1}) for $\tau$ 
to obtain a second order differential equation for
$\kappa$. 
It is clear that the Euler-Lagrange equations 
(\ref{eq:E1}) and (\ref{eq:E2}) are 
integrable; $\tau$ is given as 
a function of $\kappa$, $\kappa$ is determined as a 
quadrature. 

The Noether charge ${\cal Q}$ is identified as the 
divergence appearing in Eq.(\ref{eq:delFS}),
\begin{equation}
{\cal Q} =  f \Psi_\parallel + f_\kappa \Psi_1{}'- f_\kappa{}'\Psi_1 - 2
\tau
f_\kappa
\Psi_2\,.
\label{eq:qrig}
\end{equation}
The conserved force ${\bf F}$ is obtained by specializing the 
deformation to a constant infinitesimal translation 
$\delta {\bf  X} = {\bf  e}$ in this expression. 
In the second term, we use the Frenet-Serret
equation to obtain $\Psi_1{}' = {\bf e} \cdot {\bf n}_1{}' =
{\bf e} \cdot ( - \kappa {\bf t} + \tau {\bf n}_2 ) $.
Eq. (\ref{eq:qrig}) 
then gives 
\begin{equation}
{\bf F} = 
(f_\kappa \kappa - f ) {\bf t} +  f_\kappa{}' {\bf  n}_1
+ \tau f_\kappa {\bf  n}{}_2 
\, .
\label{eq:F1}
\end{equation}
Note that the tension in the curve is identified as 
$-F_{||}= f- f_\kappa \kappa$. It is not constant, in general, along
the curve, and may also take negative values. 

In a similar way, from the Noether charge corresponding to a rotation 
$\delta {\bf  X}= {\bf  \Omega}\times {\bf  X} $ we obtain the 
conserved torque, ${\bf M}$. In general, only terms with derivatives
of the $\Psi_i$ contribute to the differential torque ${\bf T}$. In 
this case we have $\Psi_1{}' = ( {\bf \Omega} \times {\bf X} \cdot {\bf
n}_1 ){}'
= {\bf \Omega} \cdot ( {\bf X} \times {\bf n}_1 ){}' 
= {\bf \Omega} \cdot ( {\bf t} \times {\bf n}_1 
+  {\bf X} \times {\bf n}_1{}' ).$ 
The second term contributes to the orbital part of ${\bf M}$, while 
from the first 
we find that the differential torque is given by
\begin{equation}
{\bf T}  
=  - f_\kappa\, {\bf t} \times {\bf  n}_1  = - f_\kappa \,{\bf n}_2  
\, .
\label{eq:T1}
\end{equation}
Note that while the torque due to bending, ${\bf M}$,
 is not generally of the simple form ${\bf F} \times {\bf  X}$ unless 
$f_\kappa = 0$, {\it i.e.} $f$ is constant, neither is ${\bf T}$ of the most 
general form: there is no component along either ${\bf t}$ or 
${\bf n}_1$ if only bending is penalized. This accords with our 
intuition: the axis of 
rotation due to the bending which rotates ${\bf t}$ towards ${\bf n}_1$ 
is along ${\bf n}_2$.
For this model, the second Casimir of the  euclidean group, $J$, given by 
Eq.(\ref{eq:tau}), is read off by dotting Eqs.(\ref{eq:F1}) and
(\ref{eq:T1}) as
\begin{equation}
F J=  -
 f_\kappa{}^2 \tau \,.
\label{eq:s1}
\end{equation}
We thus identify the constant appearing in Eq.(\ref{f12}) as $- F J$.

Substituting Eq.(\ref{eq:s1}) for $\tau$ into the
magnitude of the force 
 determines $\kappa$  as a quadrature, 
involving the two constants $F$ and $J$:
\begin{equation}
F^2 =  (f_\kappa{}')^2 + (f_\kappa \kappa  -f)^2   
+ { F^2 J^2\over (f_\kappa)^2}\,.
\label{eq:quad}
\end{equation}
Suppose $f_\kappa$ is not constant. The quadrature (\ref{eq:quad}) 
describes the radial motion  ($f_\kappa$) of a fictitious particle
with a mass $=2$, positive energy $F^2$, 
and angular momentum $F J$, 
moving in the central potential, 
$
V(\kappa)=  (f_\kappa \kappa - f)^2\,.
$
We note that this potential is bounded from below.
Eq.(\ref{eq:quad}) can be integrated to determine $\kappa$ implicitly as a
function of $s$:
\begin{equation}
s  = \int { d f_\kappa \over [
F^2 - V(f_\kappa)
-  F^2 J^2/ f_\kappa{}^2 ]^{1/2} }\,.
\end{equation}
Once $\kappa$, and therefore $\tau$ via Eq. (\ref{eq:s1}), is known,
one can use the expressions (\ref{t1}), (\ref{t2}),
(\ref{t3}) to obtain by a further quadrature the trajectory
in the adapted cylindrical coordinates $\{ \rho, \theta, z \}$. 
In particular, we note that from Eq. (\ref{t1}) it follows that
the radial coordinate $\rho$ is determined pointwise by the curvature
$\kappa$.

The physically most relevant model is one quadratic in $\kappa$,
subject to the constraint that the total length is constant
\cite{Lan.Sin:84}: 
\begin{equation} 
H = \int ds \, (\kappa^2 + \mu) 
\,,
\end{equation}
where $\mu$ is a Lagrange multiplier enforcing the constraint.
We have $f(\kappa) = \kappa^2 + \mu$, and Eq. (\ref{eq:quad})
reduces to
\begin{equation}
F^2 =  4(\kappa')^2  + (\kappa^2  -\mu)^2   
+ { F^2 J^2\over 4 \kappa^2}
\,,
\label{eq:quad2}
\end{equation}
where we have used $FJ = - 4 \kappa^2 \tau $.
The potential becomes  quartic for large $\kappa$. If $J$ vanishes, 
$\tau=0$, and the problem reduces 
to that of a planar elastica. There is one circular 
solution labelled by the winding number $n =\pm 1,\pm 2,
\cdots$, correponding to an $n$-fold covering of the circle, 
with $\kappa^2 = \mu$. $\mu$ is then fixed by $L$ and $n$.
The vector ${\bf F}$ vanishes on these solutions 
whereas ${\bf T}$ does not, pointing 
out of the plane.
The doubly degenerate ground states are  
the once covered circles with $n=\pm 1$.

For $n=0$, there are two oppositely oriented figure-eight configurations.
There are two inflection points 
where $\kappa=0$, connecting equal positive
and negative curvature lobes. The integrability condition,
Eq.~(\ref{eq:int}), implies
\begin{equation}
\int ds \; \kappa^2 =  \mu L\,,
\end{equation}
where $L$ is the total length of the loop --- this fixes $\mu$ to be 
positive. 

Consider next the model described by (see {\it e.g.} \cite{Gol.Lan:95})
\begin{equation}
H  = \int ds  \, (\kappa - \kappa_0)^2 \,,
\end{equation}
where $\kappa_0$ is some positive constant, the `spontaneous'
curvature (notice that we do not include a constant length constraint
in this case).
The absolute minimum, $H=0$, obtains when $\kappa=\kappa_0$, which 
corresponds to an 
$n$-fold circular loop of radius $R_0=\kappa_0^{-1}$, with $n$
arbitrary.
The ground state is therefore infinitely degenerate.
In this model we have,
\begin{equation}
F^2 = 4(\kappa{}')^2 +  (\kappa^2
-\kappa_0^2)^2 
+
{F^2 J^2\over 4 (\kappa-\kappa_0)^2} 
\,.
\end{equation}
If $J=0$, $\tau$ vanishes as well, and 
the potential possesses a single minimum at $\kappa=\kappa_0$.  
The only equilibria with
$J=0$ as before are the circles and figure eight. 
If $J \ne 0$, the potential is quite different from 
the one considered earlier: it 
now diverges at $\kappa =\kappa_0$.  
On either side of $\kappa_0$, it develops a local miminum.
As before, the integrability condition can be 
used to exclude constant $\kappa$ closed loops.
Since $F_{||} = \kappa^2 -\kappa_0^2 $, the integrability 
condition then implies that $\kappa = \kappa_0$ if it is constant.

We end this section with a brief description of the model described by 
the scale invariant bending energy with $f = \kappa$. 
In this case, Eq.(\ref{eq:E1}) implies $\tau=0$. 
We also have  ${\bf F}=0= J$.
Any plane loop extremizes this bending energy, 
which (for positive $\kappa$)
coincides with the winding number of 
the loop on this plane. The minimum is realized
on any convex loop.
\section{Torsion}
We turn now to hamiltonians of the form 
\begin{equation}
H = \int ds \; f (\tau )
\,,
\label{Hft}
\end{equation}
where $f$ is an arbitrary local function of its argument.
We will see that, like in the case $H = \int ds f(\kappa)$ discussed in the 
previous section, such models are integrable by quadratures.
We will require for the remainder of this section that neither
$\kappa$ or $\tau$ vanish.

We determine the normal variation
of the free energy ($f_\tau = \partial f / \partial \tau$):
\begin{eqnarray}
\fl
\delta_\perp H |_1 &=&
\int ds  
\left[  2 {\tau \over \kappa} f_\tau{}''
+  {\tau' \over \kappa} f_\tau{}'
- 2 {\tau \kappa'\over \kappa^2} f_\tau{}'
- f \kappa + 2  \kappa \tau f_\tau \right] \Psi_1
\nonumber \\ \fl
& & 
{}+\int ds  \left[ 2 \; {\tau \over \kappa} f_\tau \Psi_1{}' 
+ {\tau' \over \kappa} f_\tau \Psi_1
- 2 \; {\tau \over \kappa} f_\tau{}' \Psi_1 \right]'
\,, \label{eq:hp1}
\end{eqnarray}

\begin{eqnarray} 
\fl
\delta_\perp H |_2 
&=&
\int ds  
\left[ -\left( { f_\tau{}' \over \kappa} \right)''
 + {\tau^2 f_\tau{}' \over \kappa}  
- ( \kappa f_\tau  )' \right]\Psi_2
\nonumber \\ \fl
& & 
{}+\int ds  
\Big[ { f_\tau \over \kappa} \Psi_2{}''
- { f_\tau{}' \over \kappa}  \Psi_2{}'
+  ( \kappa^2 - \tau^2 ) { f_\tau \over \kappa} 
\Psi_2 
+ \left( { f_\tau{}' \over \kappa} \right)' \Psi_2
 \Big]'
\,, \label{eq:hp2}
\end{eqnarray}
where we have used Eqs. (\ref{eq:sn}), (\ref{eq:varkappa2b}).
The Euler-Lagrange  derivatives ${\cal E}_i$ along the normal 
directions  are identified as the coefficients of the $\Psi^i$
discarding total derivatives: 
\begin{eqnarray}
{\cal E}_1 
&=&
2 \tau \left( {f_\tau{}' \over \kappa} \right)'
+ {\tau' \over \kappa} f_\tau{}'
- \kappa f + 2 \kappa \tau f_\tau 
\,,
\label{eq:E21} 
\\
{\cal E}_2 
&=&
-\left( { f_\tau{}' \over \kappa} \right)''
+  {\tau^2 f_\tau{}'\over \kappa} 
- ( \kappa f_\tau )' \,.
\label{eq:E22}
\end{eqnarray}
As expected, the Euler-Lagrange equations are of order three in
derivatives of $\tau$, and therefore of sixth order in derivatives
of the embedding functions.

The  Noether charge
is given by the total derivatives in (\ref{eq:hp1}), (\ref{eq:hp2}), 
\begin{eqnarray}
\fl
{\cal Q} &=&   f \Psi_\parallel  + 2 \; {\tau \over \kappa} f_\tau
\Psi_1{}' 
- \left[ 2 \; {\tau \over \kappa} f_\tau{}'  
-
{\tau' \over \kappa} f_\tau \right]\Psi_1
+\left( { f_\tau \over \kappa} \right) \Psi_2{}''
- { f_\tau{}' \over \kappa}  \Psi_2{}'
\nonumber \\ \fl
& & {}
+ \left[\left( { f_\tau{}' \over \kappa} \right)'  + 
\left( { f_\tau \over \kappa} \right) (  \kappa^2 
+  \tau^2 )\right] \Psi_2 
\,.
\end{eqnarray}
This permits us to write down the constant force 
\begin{equation}
{\bf F} = (\tau f_\tau - f ){\bf t}
+  {\tau \over \kappa}  f_\tau{}' {\bf  n}_1 
- \left[   \left( { f_\tau{}' \over \kappa} \right)'
+ \kappa f_\tau \right] {\bf  n}{}_2
\, .
\label{eq:F2}
\end{equation}
We note that the structure of the tangential component is 
identical to the one of ${\bf F}$ for bending in Eq. (\ref{eq:F1}), 
with $ \tau \leftrightarrow \kappa$. Moreover,
the
second derivative of $\tau$ appearing in
$F_2$ can be lowered to a first derivative
by exploiting the Euler-Lagrange equation ${\cal E}_1=0$. 
Note that this was not necessary in first curvature models.
We have 
\begin{equation}
\left( { f_\tau{}' \over \kappa} \right)' + \kappa f_\tau 
= 
{1\over 2\tau} \left[ 
\kappa f - \tau' \left( { f_\tau{}' \over \kappa}
\right) 
\right]
\,, 
\end{equation}
so that we have the alternative expression:
\begin{equation}
{\bf F} 
= 
(\tau f_\tau - f ){\bf t}
+  {\tau \over \kappa}  f_\tau{}' {\bf  n}_1 
- {1\over 2\tau} 
\left[   \kappa f 
- \tau'  { f_\tau{}' \over \kappa} 
  \right] {\bf  n}{}_2\,.\label{eq:F3}
\end{equation}
The differential torque ${\bf T}$  is given by 
\begin{equation}
{\bf T} 
= - f_\tau 
{\bf  t}
 -{ f_\tau{}' \over \kappa } 
{\bf  n}{}_1 
\, .
\label{eq:t2}
\end{equation}
Notice that in addition to the expected tangential component 
$- f_\tau {\bf t}$ 
due to the twist about the `rod' axis (see {\it e.g.} \cite{Lan.Lif:70}), 
there is a contribution due to
differential twist along ${\bf n}_1$. There is, however, no ${\bf n}_2$ component. 
The second Casimir, as defined by Eq. (\ref{eq:tau}), takes the 
form
\begin{equation}
F J = f_\tau \left( f - \tau f_\tau \right) - \tau 
\left( { f_\tau{}' \over \kappa} \right)^2\,.
\label{eq:FJ2}
\end{equation}

 Remarkably the solution by quadratures 
is possible exactly as in the first curvature models, as we show 
now for a special case which is sufficient for our purposes,
\begin{equation}
f = \tau^2/ 2 + \mu\,.
\label{eq:tau2}
\end{equation} 

From Eq. (\ref{eq:F3}) we obtain  immediately
\begin{equation}
F^2 = 
{\kappa^2\over 4\tau^2} \Bigg[  \left( {\tau' \over \kappa } \right)^2 
-  {\tau^2 \over 2}  - \mu\Bigg]^2 + {(\tau^2-2\mu)^2
\over 4}
+  \tau^2 \, \left( {\tau' \over \kappa} \right)^2 \,.
\label{eq:k2m}
\end{equation}
Moreover from Eq.(\ref{eq:FJ2}), we obtain
\begin{equation}
F J   =  {1\over 2} \tau  
\left( 2\mu - \tau^2 \right) - \tau  \left( {
\tau' \over \kappa} \right)^2\,,
\label{eq:k2s}
\end{equation}
which permits to eliminate $\tau'/ k$  from
Eq.(\ref{eq:k2m}).
Doing this, we get 
\begin{equation}
\tau^4 (\tau^4 - 4\mu^2) + \kappa^2 (\tau^3 + FJ)^2 
  + 4 F J \tau^5 + 4 F^2 \tau^4 = 0\,.
\label{eq:poly}
\end{equation}
Thus $\tau$ is determined in terms of $J$, $F$ and $\kappa$ as a root
of an eighth order polynomial. 
It is rather surprising that $\tau$ is determined pointwise,  just as in
the pure bending case,  as some function of $\kappa$.

If we insist on adhering to the same mechanical analogue of 
a non-relativistic particle  with radial $\kappa$ exploited for 
bending, the 
potential appearing in the quadrature is going to be a mess. 
Fortunately, it is also 
possible to set up a quadrature for 
$\tau$. We solve  Eq.(\ref{eq:poly}) for $\kappa$ as a function of
$\tau$,
\begin{equation}
\kappa^2 = 4 \tau^4 \; { F^2 - \mu^2 + \tau FJ
+{1\over 4} \tau^4 \over ( FJ + \tau^3 )^2 }\,,
\label{eq:poly1}
\end{equation}
and 
substitute into Eq.(\ref{eq:k2s}). We obtain
an equation of the form
\begin{equation}
\tau'{}^2 + V (\tau , F , J, \mu ) = 0\,,
\end{equation}
where the potential is given by
\begin{equation}
V (\tau, F, J, \mu ) =  4 \tau^3 
\, { (F^2 - \mu^2 + \tau FJ + {1 \over 4} \tau^4 )
(\mu \tau - FJ - {1\over 2} \tau^3 ) \over
( FJ + \tau^3 )^2 }\,.
\label{eq:k2s1}
\end{equation}
which again describes a non-relativistic particle 
(this time with position $\tau$ and zero "energy") moving in a
potential which is a ratio of polynomials. 
The analysis of the equilibrium configurations
for this model is outside the scope of this paper. 
We note that the potential tends asymptotically to $\tau^4/2$ --- a
quartic once again.
Note that the integrability condition
\begin{equation}
\int ds (\tau^2 - 2 \mu ) =0 \end{equation}
implies that $\mu$ must be positive on a closed loop.

\section{Bending and torsion}
Let us now consider models with a joint dependence on $\kappa$ and $\tau$,
$f= f(\kappa , \tau )$. 
The integrability exhibited by  the cases considered so far does not
persist, in general, when $f$ depends on both $\kappa$ and $\tau$. 
Unfortunately, this is  also the case of interest in
biophysics where, for example, models of the type 
$f = \alpha \kappa^2 + \beta \tau^2$ are studied.  

The Euler-Lagrange equations take the form
\begin{eqnarray}
\label{ELktmu1}
{\cal E}_1 \, \, = \, \, 0 
&=&
f_\kappa'' 
+ (\kappa^2 -\tau^2) f_\kappa 
- \kappa f 
+ 2 \tau \left( {f_\tau' \over \kappa} \right)' 
+ \tau' {f_\tau' \over \kappa} 
+ 2 \kappa \tau f_\tau
\\
{\cal E}_2 \, \, = \, \, 0
&=&
2 (\tau f_\kappa){}' 
- \tau' f_\kappa 
- \left({f_\tau' \over \kappa} \right)'' 
+ \tau^2 {f_\tau' \over \kappa} 
- (\kappa f_\tau)'
\, .
\label{ELktmu2}
\end{eqnarray}
Note that one has to be careful to include only once the
term $-\kappa f$ in adding Eqs. (\ref{eq:E1}) and (\ref{eq:E21})
to obtain ${\cal E}_1$.
The force and the differential torque are
\begin{eqnarray}
\fl 
{\bf F} &=& (f_\tau \tau + f_\kappa \kappa - f ) {\bf t}
+ {1\over \kappa} ( \kappa f_\kappa{}' + \tau f_\tau{}' ) {\bf n}_1
- \left[   \left( { f_\tau{}' \over \kappa} \right)'
+ \kappa f_\tau  - \tau f_\kappa \right] {\bf  n}{}_2\,,
\label{eq:F12}
\\ \fl
{\bf T} &=&  - \left[  f_\tau 
{\bf  t}
 + { f_\tau{}' \over \kappa } 
{\bf  n}{}_1 +
f_\kappa {\bf n}_2 \right]\,.
\label{eq:T12}
\end{eqnarray}
The force is obtained by adding Eqs. (\ref{eq:F1}), (\ref{eq:F2}), taking
care to include the term $-f{\bf t}$ only once. The differential torque is
given by the sum of Eqs. (\ref{eq:T1}), (\ref{eq:t2}). 
Note that ${\bf T}$ has components along all directions.
It follows that the
second Casimir is
\begin{equation}
\fl
FJ = - f_\tau ( f_\tau \tau + f_\kappa \kappa - f )
- {f_\tau{}' \over \kappa^2 } ( f_\kappa{}' \kappa +
f_\tau{}' \tau ) +
f_\kappa \left[ \left({f_\tau{}' \over \kappa}
\right)' + \kappa f_\tau - \tau f_\kappa \right]\,.
\label{eq:FJ12}
\end{equation}
By comparing  this expression for $FJ$ with the corresponding one for  the model depending only
on $\tau$, it is clear that
the same strategy exploited in Sect. 5 to produce a quadrature 
will not work.
There are, however, two interesting mixed cases that are tractable by 
quadratures.  

The first possibility is to consider the  bending energy
constrained to a fixed length and torsion (see \cite{Ive.Sin:99}
for a detailed analysis of this model).
Thus, we consider the model defined by
\begin{equation}
f = {1 \over 2} \kappa^2 +  \alpha  \tau + \mu 
\,.
\end{equation}
The total torsion, ${\cal T} = \int ds \; \tau $ is dimensionless.
The fact that makes this model a minimal variation with respect to
the pure bending models of Sect. 4 is that it does not introduce 
derivatives of
$\tau$  in the equilibrium conditions.
The force and the differential torque are given by
\begin{eqnarray}
{\bf F} &=& {1\over 2} (\kappa^2 - 2\mu){\bf t}
+ \kappa' {\bf  n}_1 
+  \kappa (\tau - \alpha ) {\bf  n}{}_2\,,
\\
{\bf T}  
&=& - \alpha {\bf t} - \kappa {\bf n}_2 \,.  
\end{eqnarray}
We have then that the second Casimir is
\begin{equation}
FJ  = \alpha \mu + {\kappa^2 \over 2} ( \alpha - 2 \tau )\,,
\end{equation}
This invariant  can be inverted for 
$\tau$ as
\begin{equation}
\tau =  {1 \over \kappa^2} (  \alpha  \mu  - FJ ) + {\alpha \over 2}\,.
\label{eq:k2T}
\end{equation}
The corresponding quadrature takes then the form
\begin{equation}
\kappa'{}^2 + {1 \over 4} (\kappa^2 - 2\mu)^2 + 
{1\over 4 \kappa^2} \left( \alpha \mu - FJ + {1 \over 2} \kappa^2 \right)^2 = F^2\,.
\end{equation}
Note that the torsion constraint does not affect the integrability 
condition Eq.(\ref{eq:int}) on $F_{||}$, although it affects 
the form of the coordinates $\rho, \theta$ as defined in Sect. 3.4

The second possibility is given by adding a term linear in $\kappa$
to the model (\ref{eq:tau2}), so that
\begin{equation}
f = {1\over 2} \tau^2 + \alpha \kappa + \mu\,.
\end{equation}
The two Casimirs now take the form
\begin{eqnarray}
\fl
F^2 &=& {\kappa^2\over 4\tau^2} \Bigg[  \left( {\tau' \over \kappa } \right)^2 
-  {\tau^2 \over 2}  - \mu + \alpha {\tau^2 \over \kappa} \Bigg]^2 +
{(\tau^2-2\mu)^2
\over 4}
+  \tau^2 \, \left( {\tau' \over \kappa} \right)^2 \,,
\\ \fl
FJ &=&  {1\over 2} \tau  
\left( 2\mu - \tau^2 \right) - \left( \tau + {\alpha \kappa \over 2
\tau} \right) \left( {
\tau' \over \kappa} \right)^2
+ {\alpha \kappa \over 2 \tau} \left( \mu + {\tau^2 \over 2}
- {\alpha \tau^2 \over \kappa} \right)
\,.
\end{eqnarray}
Note that the latter is considerably more complicated
than in the pure torsion case, as given by Eq. (\ref{eq:k2s}).
It is possible to use $FJ$ to eliminate $\tau' / \kappa $ in $F^2$,
so  that $\tau$ is determined by $\kappa$ pointwise. However,
the resulting expression is quite messy.

Although it is clear that the general case will not be reducible to a
quadrature,  the use of the Casimirs of the euclidean group allows
for significant simplifications over a direct approach at
the level of the equilibrium conditions. We illustrate this fact with 
an example: let us look at the model 
\begin{equation}
f = {1 \over 2} (\kappa^2 + \tau^2 )\,. 
\end{equation}
This
is known as total curvature \cite{Str:61}, and it is a natural function
of curvature and torsion, in the sense that
${\bf n}_1{}' \cdot {\bf n}_1{}' = \kappa^2 + \tau^2$. It also appears in
Ref.
\cite{Mur.Bal:01}
as a conserved hamiltonian.
From Eqs. (\ref{ELktmu1}), (\ref{ELktmu2}), 
we read off the equilibrium conditions
\begin{eqnarray}
{\cal E}_1 &=& 2 \tau \left( {\tau' \over \kappa }\right)' 
+ {\tau'{}^2 \over \kappa } + \kappa{}'' +
{\kappa \over 2} (\kappa^2 - \tau^2 ) = 0\,, 
\label{eq:k2t2a}
\\
{\cal E}_2 &=&  - \left( \tau' \over \kappa \right)'' +
{\tau' \tau^2 \over \kappa } + \tau \kappa' = 0\,.
\label{eq:k2t2b}
\end{eqnarray}
For the force and differential torque, from Eqs. (\ref{eq:F12}), 
(\ref{eq:T12}), we obtain 
\begin{equation}
{\bf F} =
{1 \over 2} (\kappa^2+ \tau^2 ) \,{\bf t} +
{1\over 2 \kappa} (\kappa^2 + \tau^2)' \, {\bf n}_1
- \left({\tau'\over \kappa}\right)'\,{\bf n}_2\,.
\end{equation}
\begin{equation}
{\bf T} = - ( \tau {\bf t} +  {\tau'\over \kappa} \, {\bf n}_1 + \kappa
{\bf n}_2 ) 
\,.
\end{equation}
It follows that the Casimir invariants take the form
\begin{eqnarray}
F^2 
&=&
 \left[\left({\tau'\over \kappa}\right)'\right]^2 
+ {1 \over 4 \kappa^2} 
[ (\kappa^2 + \tau^2 )' ]^2  
+ {1 \over 4} (\kappa^2 + \tau^2 )^2\,.
\nonumber \\
FJ
&=&
\kappa\, \left({\tau'\over \kappa}\right)'
-  {\tau' \over 2 \kappa^2} (\kappa^2 + \tau^2 )' 
- {\tau  \over 2} (\kappa^2 + \tau^2 ) \,.
\end{eqnarray}
We can eliminate the second derivative in $F^2$ using the definition 
of $J$ to provide a condition of the form
\begin{equation}
{\cal F}( \kappa,\kappa',\tau,\tau', F, J) =0\,,
\end{equation}
which can be considered as the energy condition for the
motion of a fictitious particle in two-dimensions. 

\section{A note on perturbations} 
We have, so far, focused on exact methods in dealing with the
Euler-Lagrange  equations for the elastic models we consider. We digress
briefly in this
section to comment on the perturbative analysis of these
equations --- a more expanded treatment of this material will be
presented elsewhere~\cite{Cap.Chr.Guv:02b}. Although outside the main
focus of this article, we find it
instructive to include here an example of a complementary approach
which, through approximations, allows a complete treatment, from
the energy functional to the actual embedding that minimizes it. Apart
from the obvious benefit to intuition,
we obtain a non-trivial check of many of our formulas by
computing the force ${\bf F}$ and the Casimir $FJ$ and verifying that 
they are constant.  We do this in the particular case
\begin{equation}
f=\kappa^2 + \tau^2 + \mu^2\,,
\end{equation}
{\it i.e.}  when the total curvature is
penalized, with constrained length.  

Using Eqs. (\ref{ELktmu1}), (\ref{ELktmu2}), the equilibrium conditions 
can be satisfied, with $f$ as above, by constant (non-zero) curvature 
and torsion, 
$\kappa = \kappa_0$, $\tau=\tau_0$, provided that
\begin{equation}
\kappa_0^2 + \tau_0^2 =\mu^2
\label{k0t0}
\end{equation}
(notice that ${\cal E}_2$ is identically zero for constant curvature and
torsion). 
The resulting space curve is a circular helix. Perturbations 
would give to the axis of this helix a small curvature and torsion,
while changing, in general, $\kappa_0$ and $\tau_0$ as well. We
take then $\kappa$ and $\tau$ to be power series in a small parameter
$\epsilon$ (related to the above ``macroscopic'' curvature and
torsion of the axis of the helix) and read off the resulting
Euler-Lagrange
equations order-by-order in
$\epsilon$. The zeroth order result is Eq.~(\ref{k0t0}) above, while
to order $\epsilon$  we get
\begin{equation}
\kappa_1'' 
+ 2 \tau_0 \tau_1''
+ \kappa_1
+ \tau_0 \tau_1
=
0
\, ,
\qquad
\tau_1'''
-\tau_0 \kappa_1'
-\tau_0^2 \tau_1'
=
0
\, ,
\label{k1t1DE}
\end{equation}  
where 
\begin{equation}
\kappa(s) = \kappa_0 + \epsilon \kappa_1(s) + {\cal O}(\epsilon^2)
\, ,
\qquad
\tau(s) = \tau_0 + \epsilon \tau_1(s) + {\cal O}(\epsilon^2)
\label{ktps}
\end{equation}
and we have set $\kappa_0=1$. The solutions to~(\ref{k1t1DE})
involve constant terms, sines and cosines, as well as terms
proportional to $s$ and $s^2$. To reduce the number of parameters
(five initial conditions as well as $\mu$), we choose to eliminate the
terms in $s$ and $s^2$, resulting in the constraints
\begin{equation}
\kappa_1'(0)= (2 \mu^2 -1) \tau_0^{-1} \tau_1'(0)
\, ,
\qquad
\tau_1''(0) = \tau_0 (\kappa_1(0) + \tau_0 \tau_1(0))
\, .
\label{ss2con}
\end{equation}
This amounts to a restriction to periodic solutions, with
period equal to that of the unperturbed helix. 
The following abbreviations will be useful in this section
\[
\fl
\alpha_1 
\equiv 
\mu^{-2} 
\big( 
\tau_0 \kappa_1(0)+\alpha_2 \tau_1(0)
\big)
\, ,
\qquad
\alpha_2 
\equiv 
2 \mu^2 -1
\, ,
\qquad
\alpha_3 
\equiv 
\mu^{-2}
\big(
\kappa_1(0) + \tau_0 \tau_1(0)
\big)
\, .
\]
We may furthermore set $\tau_1'(0)=0$ by a suitable shift in $s$.
The solutions then become
\begin{equation}
\kappa_1(s)
=
 - \alpha_1 \tau_0 
+ 
\alpha_2 \alpha_3 
\cos (\mu s) 
\, ,
\qquad
\tau_1(s)
=
\alpha_1
- \alpha_3 \tau_0 
\cos(\mu s)
\, .
\label{k1t1sol}
\end{equation}
We now determine the corresponding embedding, using the Weierstrass
representation for the curve (we follow the conventions
in \cite{Lan:99}). We first solve the differential equation
\begin{equation}
\Phi'(s) = Q(s) \Phi(s)
\, ,
\end{equation}
where $\Phi(s)$ is the $SU(2)$ matrix describing the rotation of the
Frenet-Serret frame and 
\begin{equation}
Q(s) = -\tau(s) e_0 -\kappa(s) e_2
\, ,
\end{equation}
with $e_0=-\frac{i}{2} \sigma^3$, $e_1=-\frac{i}{2} \sigma^1$,
$e_2=-\frac{i}{2} \sigma^2$ and $\sigma^i$ the Pauli matrices. 
The embedding then is given by
\begin{equation}
\tilde{x}(s) + i \tilde{y}(s) = \int_0^s 2 \alpha \bar{\beta} ds'
\, ,
\qquad
\tilde{z}(s) = \int_0^s (\alpha \bar{\alpha} - \beta \bar{\beta} ) ds'
\, ,
\end{equation}
where $\alpha = \Phi_{11}$ and $\beta = \Phi_{12}$. This gives us the
helix with its tangent vector, at $s=0$, along $\hat{z}$. To get instead
its axis, at $s=0$, along $\hat{z}$, we rotate around the $x$-axis by
an angle $\eta$, with $\tan \eta = \tau_0/\kappa_0$. Denoting the
resulting embedding by $(x,y,z)$, we find
\begin{eqnarray}
\fl
x(s) 
&=&
-\mu^{-2} \cos(\mu s)
+ \epsilon  
\bigg(
\mu^{-2} \tau_0 \alpha_1 
  \big( \cos (\mu s)-1 \big)
- \frac{1}{4} \alpha_3
  \big( \cos(2 \mu s) - 1 \big)
+ \frac{1}{2} \tau_0^2 \alpha_3 s^2
\bigg)
\nonumber \\ \fl
y(s)
&=&
- \mu^{-2} \sin(\mu s)
+ \epsilon
\bigg(
\mu^{-2} \tau_0 \alpha_1 \sin (\mu s)
- \frac{1}{4} \alpha_3 \sin(2 \mu s)
+  ( \frac{1}{2} \mu \alpha_3 - \mu^{-1} \tau_0 \alpha_1)   s
\bigg)
\nonumber \\ \fl
z(s)
&=&
\mu^{-1} \tau_0  s
+ \epsilon  
\bigg( 
 \mu^{-2}(\alpha_1 + \tau_0 \alpha_3) \sin (\mu s) 
+ \mu^{-1} \tau_0 \alpha_3  \, s \cos(\mu s)
+ \mu^{-1} \alpha_1 \, s
\bigg) 
\, .
\end{eqnarray}
We notice that the axis of the helix is bent in the $x$-$z$ plane due
to the $s^2$ term in $x(s)$, while the term linear in $s$ in $y(s)$
gives it torsion as well. 
As a check of our general formulas in the previous sections, as well
as of our explicit calculations in this one, we compute now 
${\bf t}$, ${\bf n}_1$ and ${\bf n}_2$
and then the force ${\bf F}$ from~(\ref{eq:F12}). We find
\begin{equation}
{\bf F}= 
-2 \epsilon \mu^3  \alpha_3  \, {\bf y}
\, ,
\label{Fsol}
\end{equation}
showing that the (indeed constant) force only appears as a result of 
the perturbation,
an exclusive feature of this particular model. Finally,
Eq.~(\ref{eq:FJ12}) gives that $FJ$ is zero to this order in $\epsilon$,
$FJ=0+ {\cal O}(\epsilon^2)$.

A final remark is due concerning the validity of the above first-order
solution. When moving on to second order, the solutions for
$\kappa_2$, $\tau_2$, will again
involve periodic as well as non-periodic terms. Eliminating the latter
fixes some of the parameters that are arbitrary in the first-order
results. More generally, requiring periodicity at order $k$, restricts
the solutions found to orders less than $k$, 
a feature that can be traced to the non-linearity of the Euler-Lagrange
equations. 
\section{Filament model recursion scheme}
\label{FM}
In this section, we discuss briefly the filament model recursion scheme,
and its relationship with the Noether currents for curves.
It is described in great detail by Langer
in the nice review \cite{Lan:99}, and our brief discussion is contained
in his work.
We are only changing the point of
view by putting hamiltonians to the forefront, rather than
curve motions. 

As we mentioned
briefly at the end of Sect. 2,  consider spatial vector fields
${\bf Y}$ which locally preserve arclength, then the filament model 
recursion
scheme is defined by
\begin{equation}
{\bf t} \times {\bf Y}_{(n)} = {\bf Y}_{(n-1)}{}'\,,     
\quad \quad \mbox{with} \quad {\bf Y}_{(0)} = - {\bf t}\,.
\end{equation}
The first few terms in 
this hierarchy are
\begin{eqnarray}
{\bf Y}_{(1)} &=& \kappa {\bf n}_2\,,
\nonumber \\
{\bf Y}_{(2)} &=&  {\kappa^2 \over 2} {\bf t}
+ \kappa' {\bf n}_1 + \kappa \tau {\bf n}_2\,,
\nonumber \\
{\bf Y}_{(3)} &=& \kappa^2 \tau {\bf t} + (2 \tau \kappa' 
+ \kappa \tau' ) {\bf n}_1 + (\kappa \tau^2 - \kappa{}'' - {\kappa^3 \over
2} ) {\bf n}_2\,.
\nonumber 
\end{eqnarray}
These vector fields have remarkable properties. 
First, we recognize that ${\bf Y}_{(1)}$, known as the filament model, is
the differential
torque ${\bf T}$ for the model $f = (1/2) \kappa^2$,  and ${\bf Y}_{(2)}$
is both the force ${\bf F}$ for the same model and also the differential
torque for the model $f = \kappa^2 \tau$.  Moreover, the
integral of the tangential 
component of the $n$-th gives the corresponding conserved hamiltonian.

Now, is it possible to set up alternative recursion schemes?
From the point of view of the hamiltonians, one can start start with some
$H = \int ds \; f (\kappa , \tau )$, and compute its differential torque
${\bf T}$
set ${\bf T} = {\bf Z}_{(1)}$, then from Eq. (\ref{eq:TFt}) it follows
that the associated force is ${\bf F} = {\bf Z}_{(2)}$. Now from the 
equilibrium conditions in the form (\ref{eq:Eq}), we have
$ {\cal E}^i {\bf n}_i = {\bf t} \times {\bf Z}_{(3)}$, so that
${\bf Z}_{(3)} = Z_{(3)\,\parallel} {\bf t} 
+ {\cal E}_2 {\bf n}_1 -    {\cal E}_1 {\bf n}_2 $.
However, to satisfy the condition that the vector be arclength preserving,
we
need to satisfy $Z_\parallel{}' - \kappa {\cal E}_2 = 0$,
and this `perfect derivative phenomenon' appears to happen only in the 
filament model recursion scheme. For example, for the model quadratic in
$\tau$, $\kappa {\cal E}_2$ is not a total derivative.
   
\ack
R C was supported by
CONACyT grant 32187-E. C C and J G were supported by
CONACyT grant 32307-E and DGAPA-UNAM grant IN119792.

\end{document}